\documentclass[twocolumn,showpacs,superscriptaddress,prl]{revtex4}%
\pdfoutput=1
\usepackage{amsmath}
\usepackage{amsfonts}
\usepackage{graphicx}
\usepackage{longtable}
\usepackage{amssymb}%
\usepackage[toc,page]{appendix}
\providecommand{\U}[1]{\protect\rule{.1in}{.1in}}

\newcommand{\qed}{{\hfill$\Box$}}

\begin{document}
\title{Uncertainty Relations for General Unitary Operators}
 
\author{Shrobona Bagchi}
\email{shrobona@hri.res.in}
\affiliation{
Harish-Chandra Research Institute, Chhatnag Road, Jhunsi, Allahabad 211 019, India and\\ and Homi Bhabha National Institute, BARC Training School Complex, Anushakti Nagar, Mumbai 400 085, India}
\author{Arun Kumar Pati}
\email{akpati@hri.res.in}
\affiliation{
Harish-Chandra Research Institute, Chhatnag Road, Jhunsi, Allahabad 211 019, India and\\ and Homi Bhabha National Institute, BARC Training School Complex, Anushakti Nagar, Mumbai 400 085, India}


\date{\today}

\begin{abstract}
We derive several uncertainty relations for two arbitrary unitary operators acting on physical states of a Hilbert space. 
We show that our bounds are tighter in various cases than the ones existing in the current literature. Using the uncertainty relation for the unitary operators we obtain the tight state-independent lower
bound for the uncertainty of two Pauli observables and anticommuting observables in higher dimensions.
With regard to the minimum uncertainty states, 
we derive the minimum uncertainty state equation by \textit{analytic method}, and relate this to the ground-state problem of the Harper Hamiltonian.
Furthermore, the higher dimensional limit of the uncertainty relations and minimum uncertainty states are explored. 
From an operational point of view, we show that the uncertainty in the unitary operator is directly related to the visibility
of quantum interference in an interferometer where one arm of the interferometer is affected by a unitary operator. This shows
a principle of preparation uncertainty, i.e., for any quantum system, the amount of visibility for two general non-commuting unitary operators is non-trivially upper bounded.

\end{abstract}

\maketitle

\section{ I. Introduction}

Undoubtedly, the Heisenberg uncertainty principle is one of the fundamental concepts in
quantum theory \cite{Heisenberg}. In early development of quantum theory this played a pivotal role 
which provided deep insights into the nature of quantum world and the way this differs from
the classical world. The uncertainty principle sets limits on our ability to predict the outcomes
of two incompatible measurements, and it was originally formulated by
Heisenberg~\cite{Heisenberg} for position and momentum observables. However, a rigorous proof of
the uncertainty relation was presented by 
Robertson~\cite{rob} and Schr\"{o}dinger~\cite{sch} for 
arbitrary pairs of non-commuting observables $A$ and $B$. This is given by
\begin{equation}
\Delta A \Delta B \geq\frac{1}{2}\,|\langle \psi | [A,B]| \psi \rangle|,
\end{equation}
where the uncertainties in the observables are quantified in terms of the
standard deviations $\Delta A=\sqrt{\langle A^{2}\rangle-\langle
A\rangle^{2}}$, $ \Delta B=\sqrt{\langle B^{2}\rangle-\langle
B\rangle^{2}}$ and $\langle O \rangle = \langle \psi|O|\psi \rangle$ is the average for an observable $O$ 
in the state $|\psi \rangle$. Recently, stronger uncertainty relations are proved which go
beyond the Robertson uncertainty relations and they capture the notion of incompatible 
observables \cite{Maccone}. Uncertainty relations are at the center stage of current research in quantum theory and quantum information \cite{Weigert, Werner,Abbott, Yao}. 
The uncertainty relations are useful from the point of view of foundational aspects  and have applications in quantum technology as well
\cite{Busch1, Lahti, Hall, Guhne, Hofmann, Peres, Huang}.

In quantum theory, the linear superposition principle gives rise to quantum interference and the uncertainty  
in physical observables. The same notion gives rise to the entangled states in composite quantum systems. There exists both the preparation uncertainty relations as well as the prominent measurement disturbance relations that collectively capture the uncertainty associated with incompatible observables. Measurement disturbance relations have long played an important role in distinguishing the classical world from the quantum world. Among the various preparational uncertainty relations and the measurement disturbance relations,
note that the Robertson uncertainty relation is 
not about measurement disturbance, rather it is about preparation uncertainty for a quantum system. This entails that we cannot 
prepare an ensemble of quantum systems for which two non-commuting observables will have arbitrary uncertainty.
Now, one may ask, can the preparation uncertainty be seen in the quantum interference experiment. Can one 
formulate an uncertainty relation that is directly testable in interference? We will show that the 
uncertainty relation for two arbitrary unitary operators can reveal the preparation uncertainty. In this
sense, these uncertainty relations will unify two fundamental features of quantum world, namely, the 
interference and the uncertainty.

Interestingly, Massar and Spindel have proved an uncertainty relation for two unitary operators that obey the commutation 
relation $UV = e^{i \phi} VU$ which also applies for the discrete Fourier transform \cite{Massar}.
This tells us the extent to which a quantum state can be simultaneously localized in two mutually unbiased bases that 
are related by a discrete Fourier transform. 
In addition, this relation can interpolate between finite dimensional system to continuous variable
cases. 
However, the relation obtained by Massar-Spindel \cite{Massar} is not tight for higher dimensional systems. A few other results can be found in Ref. \cite{Marcelo1, Marcelo2, Soto, namiki},
where the authors have discussed the uncertainty lower bounds for the unitary operators related by the discrete Fourier transform.
However, there are no uncertainty relation for general unitary operators.  

In this paper, we show that the preparation uncertainty can be captured via the uncertainty relations for unitary operators.
We derive several uncertainty relations for two arbitrary unitary operators which are stronger than
those existing in the current literature. Also, we show that our uncertainty relation can be used to derive tight state independent bound for two arbitrary Pauli observables
and two anticommuting observables in higher dimensions. With regard to the minimum uncertainty states, 
we derive the  equation of the critical states for the product of the unitary operators using the \textit{Jackiw analytical method} \cite{Jackiw}. We connect this minimum uncertainty state equation
equation for two unitary operators related by the discrete Fourier transform to the problem of finding the ground state of the Harper Hamiltonian \cite{Harper}. 
We also show that in the infinite-dimensional limit, one recovers 
the minimum uncertainty state equation of the canonical observables from the minimum uncertainty state equation of the two discrete unitary operators related by the 
discrete Fourier transform. Again, we show that the uncertainty relations for the unitary operators reduce to the well known uncertainty relations 
for the Hermitian operators for some set of states and unitary operators.

The paper is organized as follows. In section II, we define the uncertainty quantifiers for the unitary operators. Here, we introduce and 
discuss the operational significance of the uncertainty relations of unitary operators
in terms of state preparation uncertainty and interference visibility. In section III, we provide uncertainty relations for two general unitary operators. 
In section IV, we present the unitary operators which obey the generalized Clifford algebra and some of their important properties, which will be used in our analysis.
In section V, we briefly discuss the minimum uncertainty states of the unitary operators in general.  
In section VI, we discuss the higher dimensional limit of the uncertainty relations and the minimum uncertainty state equation. We end with  discussions and conclusion in section VII.

\section{ II. The Uncertainty quantifiers}

In the existing literature on the uncertainty relations, various measures of the uncertainty have been proposed. One such measure of the uncertainty is the variance. 
The variance based uncertainty relation for the unitary operators are defined analogously to that of the canonical observables in the following manner.
Let $U$ and $V$ be any two arbitrary finite dimensional unitary operators. The uncertainties associated with $U$ and $V$ in the state $|\psi \rangle$ are defined as \cite{Massar}
\begin{align}
\Delta U^2  &= \langle \psi|U^{\dagger}U |\psi \rangle - \langle \psi|U^{\dagger}|\psi \rangle \langle \psi|U |\psi \rangle  \nonumber\\
&= 1- |\langle \psi|U|\psi \rangle|^2 \nonumber\\
\Delta V^2  &= \langle \psi|V^{\dagger}V |\psi \rangle - \langle \psi|V^{\dagger}|\psi \rangle \langle \psi|V |\psi \rangle  \nonumber\\
&= 1- |\langle \psi|V|\psi \rangle|^2 
\end{align}
with $0 \le \Delta U^2  \le 1 $ and $0 \le \Delta V^2  \le 1 $. The uncertainties in any unitary operator has a clear physical meaning. It is related 
to the Fubini-Study metric \cite{Fubini,Study} on the projective Hilbert space ${\cal P(H)}$ of the quantum system. The Fubini-Study metric for two quantum states 
(say) $|\psi_1 \rangle$ and $|\psi_2 \rangle$ is defined as \cite{Anandan, Pati1}
\begin{align}
S(\psi_1, \psi_2)^2 = 4( 1- |\langle \psi_1|\psi_2 \rangle|^2 ).
\end{align}
If we define $\vert\psi_1\rangle=U\vert\psi\rangle$ and $\vert\psi_2\rangle=V\vert\psi\rangle$, then the uncertainty in any unitary operator is nothing but the distance between the original and the 
unitarily evolved quantum state (up to a constant
factor). The uncertainty relation for two non-commuting operators then limits how well
we can distinguish two different unitary evolutions of a state from the original
state. It turns
out that quantum theory has an intrinsic preparation uncertainty, i.e.,  we cannot prepare a quantum state for which the sum of two distinguishable metrics will be arbitrarily small.

There is another operational significance of the uncertainty for unitary operators. The interference fringe visibility can be linked to the uncertainty of the unitary operators. 
If we send a particle in a pure state $|\psi\rangle$ through a Mach-Zhender interferometer
and apply a unitary operator in one arm of the interferometer, then the visibility ${\cal V}$ is governed by ${\cal V}= |\langle \psi|U|\psi \rangle|$. Thus, we have the relation 
${\cal V}^2 + \Delta U^2 =1$ \cite{sjoqvist}. This shows a strict complementarity between the interference visibility and the uncertainty in the unitary operator. Therefore, any restriction on 
the uncertainty in the unitary operator will place a restriction on the interference visibility. The uncertainty relations for two non-commuting operators highlights a 
preparation uncertainty similar to that of the Robertson-Schr{\"o}dinger uncertainty relation. This says that quantum states cannot be prepared which will display maximum visibility 
in the interference for two non-commuting unitary operators. Hence, the amount of visibility for two general unitary operators is non-trivially upper bounded.

\section{ III. uncertainty relations for two arbitrary unitary operators}

Various techniques have been developed in the past for analyzing the uncertainty relations of the Hermitian operators. 
Many of them are state-dependent uncertainty relations \cite{sch, Maccone}. However, they have not been applied in the case of unitary operators yet. 
Thus, using some of the important techniques explored for the case of uncertainty relation of the Hermitian operators, we derive the following uncertainty relations for 
the general unitary operators. In several occasions, these bound give better bounds than the existing bounds for the variance based uncertainty relations for the unitary operators.
\vskip 5pt
{\bf Uncertainty relation-1:} The sum of uncertainties in two unitary operators $U$ and $V$ are lower bounded as
\begin{align}
\Delta U^2  + \Delta V^2 &\geq 1+ \vert \langle\Psi_U \vert\Psi_V\rangle\vert^2 -2\cos\Phi |\Delta^{(3)}|,
\end{align}
where $\Phi= Arg \Delta^{(3)}$ and $\Delta^{(3)}$ is the three-point Bargmann invariant \cite{Bargmann} defined as 
$ \Delta^{(3)}=  \langle\Psi| \Psi_U\rangle \langle\Psi_U| \Psi_V\rangle \langle\Psi_V| \Psi \rangle$ with $|\Psi_U\rangle= U|\Psi\rangle$ and $|\Psi_U\rangle= U|\Psi\rangle$. 
\vskip 5pt
To prove this, let us define two vectors in Hilbert space: 
\begin{align}
\vert\Psi_1\rangle=(U-\langle U\rangle)\vert\Psi\rangle,~~\vert\Psi_2\rangle=(V-\langle V\rangle)\vert\Psi\rangle.
\end{align}
From the above definitions, we have $\langle \Psi_1\vert\Psi_1\rangle= \Delta U^2$ and $\langle \Psi_2\vert\Psi_2\rangle= \Delta V^2$. Now,
$\langle\Psi_1\vert\Psi_2\rangle=\langle\Psi\vert U^{\dagger}V\vert\Psi\rangle -\langle V\rangle\langle U^{\dagger}\rangle$. Using this we get 
\begin{align}\nonumber
\vert\langle\Psi_1\vert\Psi_2\rangle\vert ^2=(1-\Delta U^2 )(1-\Delta V^2 )+ \vert\langle\Psi\vert U^{\dagger}V\vert\Psi\rangle\vert^2\\ 
-\langle\Psi\vert U^{\dagger}V\vert\Psi\rangle\langle U\rangle\langle V^{\dagger}\rangle-\langle V\rangle\langle U^{\dagger}\rangle\langle\Psi\vert V^{\dagger}U\vert\Psi\rangle.
\end{align}
Now, we apply the Cauchy-Schwartz inequality, i.e., $\vert\langle\Psi_1\vert\Psi_2\rangle\vert^2\leq \Delta U^2\Delta V^2$ and this leads to:
\begin{align}\nonumber
&\Delta U^2  + \Delta V^2  \geq \vert\langle\Psi\vert U^{\dagger}V\vert\Psi\rangle\vert^2 +1 \\ 
& -\langle\Psi\vert U^{\dagger}V\vert\Psi\rangle\langle U\rangle\langle V^{\dagger}\rangle-\langle V\rangle\langle U^{\dagger}\rangle\langle\Psi\vert V^{\dagger}U\vert\Psi\rangle.
\end{align}
The above equation can be written compactly using the three-point Bargmann invariant
$ \Delta^{(3)}$, where $\Delta^{(3)}=  \langle\Psi| \Psi_U\rangle \langle\Psi_U| \Psi_V\rangle \langle\Psi_V| \Psi \rangle$  and $Arg \Delta^{(3)} = \Phi$. With these, we can express Eq(7) as
\begin{align}
\Delta U^2  + \Delta V^2 &\geq 1+ \vert \langle\Psi_U \vert\Psi_V\rangle\vert^2 -2\cos\Phi |\Delta^{(3)}|
 \end{align}
However, $|\Delta^{(3)}|= \sqrt{(1-\Delta U^2)(1-\Delta V^2)}\vert\langle\Psi_U\vert\Psi_V\rangle\vert$. 
This gives the following equation
\begin{align}\nonumber
 \Delta U^2 & + \Delta V^2  \geq 1 + \vert\langle\Psi_U\vert\Psi_V\rangle\vert^2 \\ 
 &-2\cos\Phi\sqrt{(1-\Delta U^2)(1-\Delta V^2)} \vert\langle\Psi\vert U^{\dagger}V\vert\Psi\rangle\vert.
\end{align}
Hence the proof. The above relation can be generalized for mixed states also as given in Appendix A using an alternate derivation. Recently, uncertainty relation-1 in Eq(4) is derived using the product representation formula for the weak values \cite{Hall3}.

Unlike the product of the uncertainties, the sum of uncertainties does not face the triviality problem. Since the uncertainty relation is a sum of variances, 
the lower bound is not only dependent on the incompatibility of two unitary operators but also on the incompatibility of the unitary operator on the specific states. If the operators do not share 
a common eigenstate, there is a non-zero amount of uncertainty shown by the lower bound if the state is eigenstate of one of the operators. 

The terms $\vert \langle \psi_U|\psi_V\rangle\vert$ and $|\Delta^{(3)}|$ reveal the incompatibility
of the two unitary operators and the incompatibility of the unitary operators on the specific states. 
To see this, note that if there are no joint eigenstates of $U$ and $V$, then for the eigenstate of either of $U$ or $V$, 
the above terms reveal a remnant uncertainty in the lower bound. 
In such cases, the term $\vert \langle \psi_U|\psi_V\rangle\vert$ is equal to one when the condition $V\vert\Psi\rangle=e^{i\phi}U\vert\Psi\rangle\nonumber$
is satisfied. In other cases, if there exists at least one joint eigenstate of $U$ and $V$, then the uncertainty relation is trivially satisfied with equality. 

A weaker version of uncertainty relation-1 can be obtained which reads as
\begin{equation}
 \Delta U^2  + \Delta V^2  \geq 1 + \frac{|\langle \psi_U|\psi_V \rangle |^2 -  |\langle \psi_U|\psi_V \rangle ||\cos\Phi| }{1- |\langle \psi_U|\psi_V \rangle ||\cos \Phi|}.
\end{equation}
This equation is the consequence of the relation $\sqrt{(1-\Delta U^2)(1-\Delta V^2)}\leq 1-\big(\frac{\Delta U^2+\Delta V^2}{2}\big)$.
Another weaker version of the uncertainty relation can be found which is given by 
\begin{align}\Delta U^2+\Delta V^2\geq 1-\vert\langle U^\dagger V\rangle\vert.\end{align} 
This equation follows from Eq(7) by eliminating $|\cos\Phi|$ and using the fact that $|\cos\Phi|\geq 0$.
It tells that higher is the joint visibility term $\vert \langle \psi_U|\psi_V\rangle\vert$, lower is the uncertainty bound for the unitary operators.

The term $\cos\Phi$ characterizes the phase difference between phases of $\langle U^\dagger V\rangle$ and $\langle U^\dagger\rangle\langle V\rangle$ \cite{Pancharatnam, Oi}.
Note that $\cos\Phi=1$, i.e., $\Phi=0$ when the state is either the eigenstate of $U$ or $V$ or $U^\dagger V$ or when $Cov(U,V)=0$, i.e., $\langle U^\dagger V\rangle=\langle U^\dagger\rangle\langle V\rangle$, 
even when the operators $U$ and $V$ are incompatible. Thus it characterizes the difference in phase shifts. To see the dependence of the uncertainty lower bound on $\cos\Phi$, first note that we can write 
$|\langle U^\dagger V\rangle|=\sqrt{1-\Delta(U^\dagger V)^2}$. Using this in Eq(8) we obtain the following equivalent form
\begin{align} 
\Delta U^2+\Delta V^2\geq 2-\frac{\Delta (U^\dagger V)^2}{1-\sqrt{1-\Delta(U^\dagger V)^2}|\cos\Phi|}.
\end{align}
It shows that lower is the difference between the phase shifts, smaller is the value of the uncertainty lower bound.

Now, we characterize the states that saturate the equality of the above uncertainty relation. 
The necessary condition for equality is the Cauchy-Schwartz equality condition.
The states that satisfy the Cauchy-Schwartz equality condition have the form $\vert \Psi_1\rangle \propto \vert \Psi_2\rangle$,
i.e., $(U-\langle U\rangle)\vert\Psi\rangle \propto (V-\langle V\rangle)\vert\Psi\rangle$. 
A priori it is not evident whether the minimum uncertainty states saturate the Cauchy-Schwartz equality condition. 
Therefore, we check this in dimensions $d\geq 2$ 
for the discrete unitary operators related by the discrete Fourier transform. 
From Ref.\cite{Massar}, one may note that the minimum uncertainty states of such unitary operators are the ground states of the 
Harper Hamiltonian \cite{Opatrny, Harper}, which will be discussed in more detail in
section V below.
We have taken the ground states of the Harper Hamiltonian for dimensions $3,~5,~8,~12$, and found that our bound is not saturated for $d\geq 3$.
It is clear from the Table I that though the uncertainty relations are not tight, but they are tighter than that given in Ref.\cite{Massar}. 
Later, we prove that the bound given by the uncertainty relation-1 is tight for the unitary operators related by the discrete Fourier transform in the infinite 
dimensional limit, i.e., for the phase space translation operators in infinite-dimensional Hilbert space, we show that our bound is saturated asymptotically for the above said class of unitary operators. 
For an arbitrary state, we have plotted our bound in Fig[1], which shows that our bound performs better than the Massar-Spindel bound.

\begin{figure}
\centering
\includegraphics[width=0.23\textwidth]
{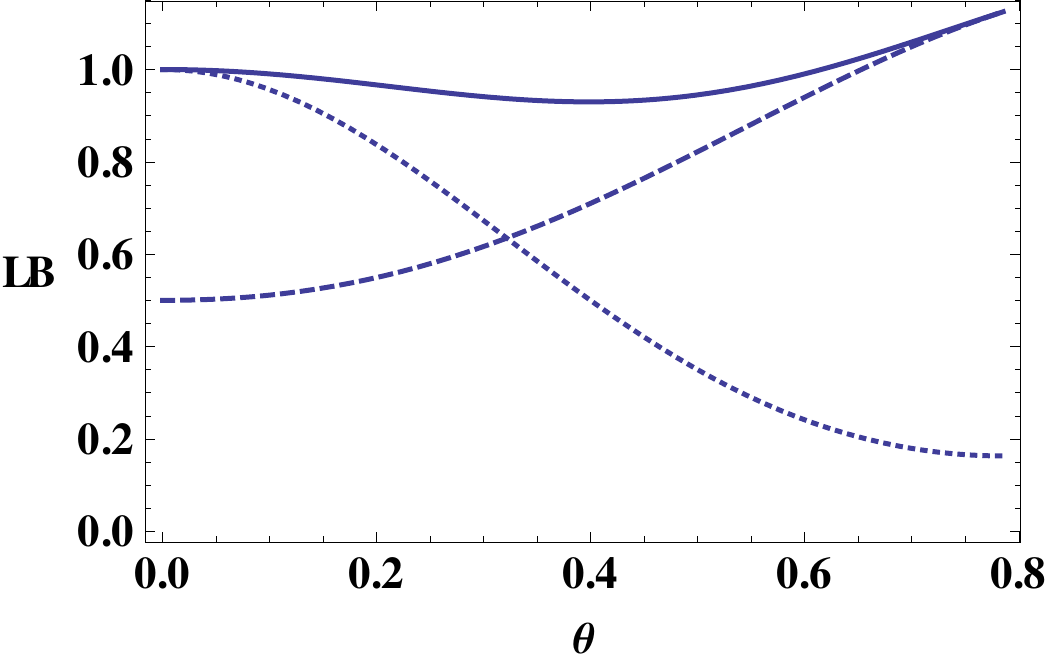}
\includegraphics[width=0.242\textwidth]
{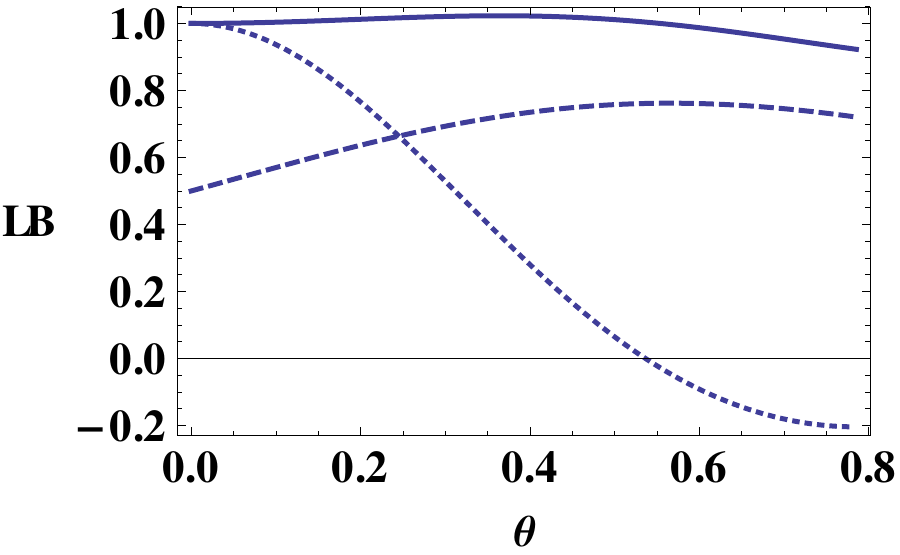}
\includegraphics[width=0.242\textwidth]
{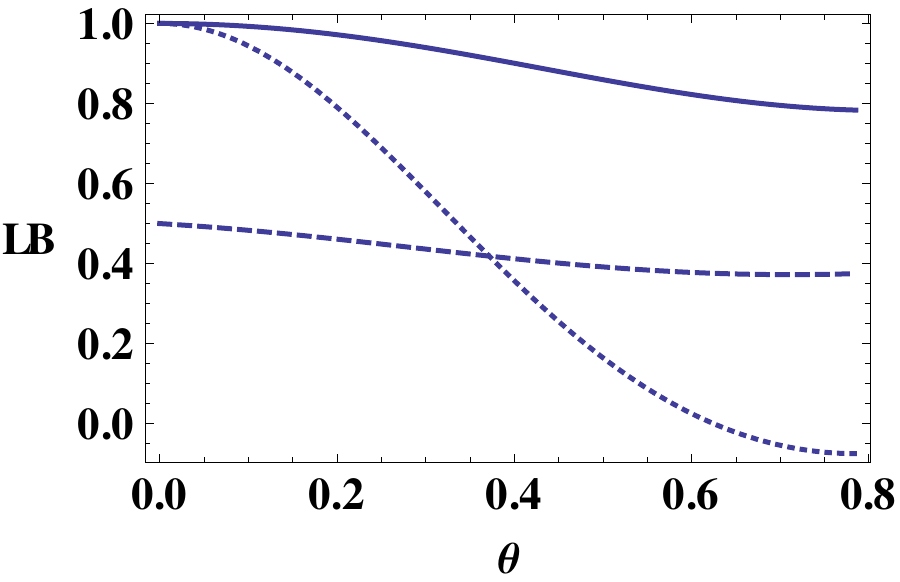}
\includegraphics[width=0.23\textwidth]
{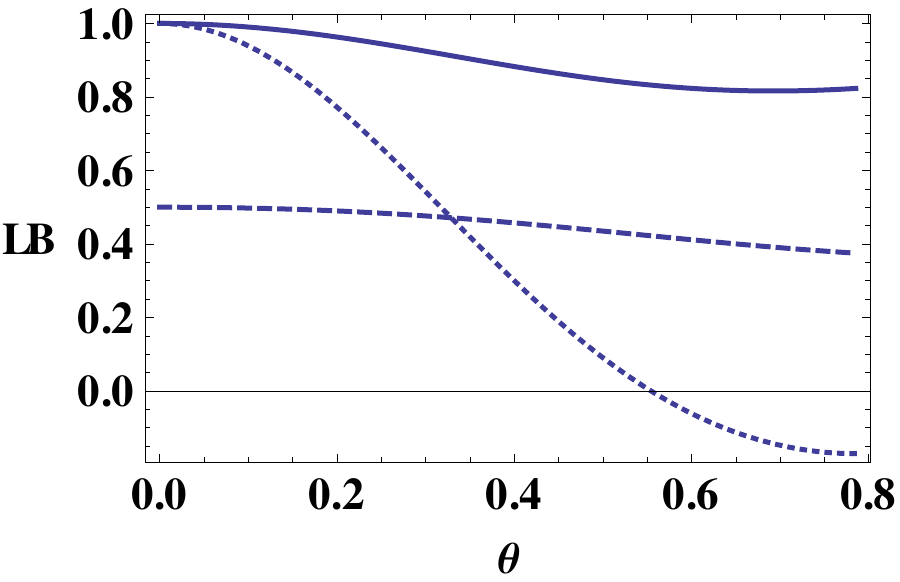}
\caption{\footnotesize{(Colour online) Lower bounds of the three different bounds given by uncertainty relation of Massar-Spindel (dotted), uncertainty relation-1 (solid), and uncertainty relation
3 (dashed)
plotted together for the three, five, eight, and twelve (starting from top left, in clockwise direction) dimensional unitary operators related by the discrete Fourier 
transform. We have shown the bounds for the class of states parametrized as $\vert\Psi\rangle=cos\theta\vert 0\rangle -sin\theta\vert d-1\rangle$, $d$ being the dimension of the Hilbert space, 
in the range
$0\leq \theta\leq \frac{\pi}{4}$. In this region, we have $\cos\Phi\geq0.$}}
\end{figure}
\vskip 5pt
\subsection{Application-1}
\subsubsection{Tight state-independent uncertainty lower bound for a pair of qubit observables}
Here, we show that one can obtain the tight state-independent uncertainty relation for two arbitrarty Pauli observables \cite{Abbott} from the uncertainty relation-1.
Let us consider $U=\bold{a}.\bold{\sigma}$ and $V=\bold{b}.\bold{\sigma}$ to be a pair of Pauli observables which are both 
unitary and Hermitian, simultaneously. Here, $\bold{a}$ and $ \bold{b}$ are unit vectors.
Then, we have 
\begin{align}
\vert\langle U^\dagger V\rangle\vert^2 = \vert\langle UV\rangle\vert^2=
\vert\langle(\bold{a}.\bold{\sigma})(\bold{b}.\bold{\sigma})\rangle\vert^2 
\end{align}
Since $U$ and $V$ are Pauli observables, therefore $\bold{a}, ~\bold{b}$ are real vectors. Let, $\rho=\frac{1}{2}(I+\bold{r}.\bold{\sigma})$, where
$ \bold{r}$ is a real vector. With this, we have the following relation
\begin{align}
\vert\langle U^\dagger V\rangle\vert^2 = \vert\langle UV\rangle\vert^2= \vert(\bold{a}.\bold{b})\vert^2 + \vert ((\bold{a}\times\bold{b}).
\bold{r}) \vert^2.
\end{align}
The uncertainty relation-1 for qubits in
case of the Pauli observables can be expressed as
\begin{align}\nonumber
\Delta U^2 + \Delta V^2  \geq 1 + \vert\bold{a}.\bold{b}\vert^2-\\   2\vert \bold{a}.\bold{b}\vert\sqrt{1-\Delta U^2}\sqrt{1-\Delta V^2}+
\vert((\bold{a}\times\bold{b}).\bold{r})\vert^2.
\end{align}
Now the last term of the the equation  which is state dependent is always positive, therefore we have
\begin{align}\nonumber
  \Delta U^2 + \Delta V^2  \geq 1 + \vert\bold{a}.\bold{b}\vert^2-\\ 
  2\vert \bold{a}.\bold{b}\vert\sqrt{(1-\Delta U^2)(1-\Delta V^2)}.
\end{align}
This is nothing but the tight state-independent uncertainty relation for Pauli observables. Hence the proof. Note that if we want state-dependent uncertainty relation, then Eq(15) is tighter than Eq(16).
\vskip 5pt
\subsection{Application-2}
\subsubsection{Tight state-independent uncertainty lower bound for two anticommuting observables in higher dimensions}
The uncertainty relation for the unitary operators can be invoked to obtain the tight state-independent uncertainty relation for the anticommuting observables in higher dimensions of the form $2^n$ for any positive integer $n$.
We can express the anticommuting observables in any $2^n$ dimension as $\Gamma_a= \bold{a}.\bold{\Gamma}$ and $\Gamma_b=\bold{b}.\bold{\Gamma}$ \cite{Winter},
where $\bold{a}$ and $\bold{b}$ are real vectors in any finite dimensional vector space. Here, $\bold{\Gamma}=\{\Gamma_1, \Gamma_2...\Gamma_{2n}\}$, 
where $\Gamma_i$ are the generators of the Clifford algebra. It can be easily deduced that the above mentioned anticommuting observables follow 
the following state-independent uncertainty relation
\begin{align}\nonumber
 \Delta \Gamma_a^2+\Delta \Gamma_b^2\geq&1+|\bold{a}.\bold{b}|^2-\\ 
 &2|\bold{a}.\bold{b}|\sqrt{(1-\Delta \Gamma_a^2)(1-\Delta \Gamma_b^2)}.
\end{align}
This uncertainty relation is tight in the sense, that we can always find  a quantum state that saturates this bound. The form of the state is same as before given in \cite{Abbott},
but only in terms of the $\Gamma_i$ as $\rho=\frac{1}{d}(I+\bold{g}.\bold{\Gamma})$, where $\bold{g}_{\pm}=
\sqrt{1-\Delta\Gamma_a^2}\bold{a}\pm\tau\frac{\Delta\Gamma_a}{1-(\bold{a}.\bold{b})^2}(\bold{b}-(\bold{a}.\bold{b})\bold{a})$ is a pure state 
and $\tau=sgn(\bold{a}.\bold{b})$ \cite{Abbott}. This shows that the uncertainty relation-1 has interesting implications for qubits and higher dimensional systems.
\vskip 5pt
{\bf Uncertainty relation-2:} The sum of uncertainties of two unitary operators $U$ and $V$ are lower bounded
\begin{align}
\Delta U^2 + \Delta V^2 \geq \vert\langle\Psi\vert U^\dagger\pm iV^\dagger\vert \Psi^\perp\rangle\vert^2 
\mp 2Im[Cov(U,V)],
\end{align}
where $ Cov(U,V)= \langle U^\dagger V\rangle-\langle U^\dagger\rangle\langle V\rangle$.
\vskip 5pt
To prove this, let us define two operators $C=U-\langle U\rangle$ and $D=V-\langle V\rangle$. Then, we have  
$\Delta U=\vert\vert C\vert\Psi\rangle\vert\vert$ and $\Delta V=\vert\vert iD\vert\Psi\rangle\vert\vert$. Note that  
\begin{align} 
\vert\vert (C+iD)\vert\Psi\rangle\vert\vert^2=
\Delta U^2+\Delta V^2-
2Im[Cov(U,V)]. 
\end{align}
Now, by the Cauchy-Schwartz inequality we have, 
$\langle\Psi\vert(C^\dagger -iD^\dagger)( C+iD)\vert\Psi\rangle\geq \vert\langle\Psi\vert C^\dagger-iD^\dagger\vert\Psi^{\perp}\rangle\vert^2$. However, 
$\vert\langle\Psi\vert C^\dagger-iD^\dagger\vert\Psi^{\perp}\rangle\vert^2= \vert\langle\Psi\vert U^\dagger-iV^\dagger\vert\Psi^{\perp}\rangle\vert^2$. 
Combining these equations, we can obtain one of the above inequality.
Again, by starting from the condition 
$ \vert\vert (C-iD)\vert\Psi\rangle\vert\vert^2=
\Delta U^2+\Delta V^2+
2Im[Cov(U^\dagger V)]$, and following exactly same steps, we obtain, the other inequality in the above equation.
Hence the proof. 

It is easy to see that the term $\vert\langle\Psi\vert U^\dagger\pm iV^\dagger\vert \Psi^\perp\rangle\vert^2 $ expresses the incompatibility of the two unitary operators on the states. 
This is because, if the two unitary operators
do not share a common eigenstate, and unless $|\Psi\rangle$ is an eigenstate of $U\pm iV$, the eigenstate of either $U^\dagger$ or $V^\dagger$ is not an eigenstate of $U^\dagger\pm iV^\dagger$ and thus a remnant
uncertainty is shown in the lower bound for such a case.
We have plotted this lower bound for the discrete unitary operators related by discrete Fourier transform in the dimensions $3,~5,~8$ and $12$. A comparison has been made with 
the other bounds. The plot show that this bound also performs better than the Massar-Spindel bound \cite{Massar} for certain regions in the state space.
The equality condition is same as the Cauchy-Schwartz equality condition. To check the bound for the minimum uncertainty states, we have shown in Table I that this uncertainty relation very closely 
saturates the minimum uncertainty states.
\begin{figure}[t]
\centering
\includegraphics[width=0.23\textwidth]
{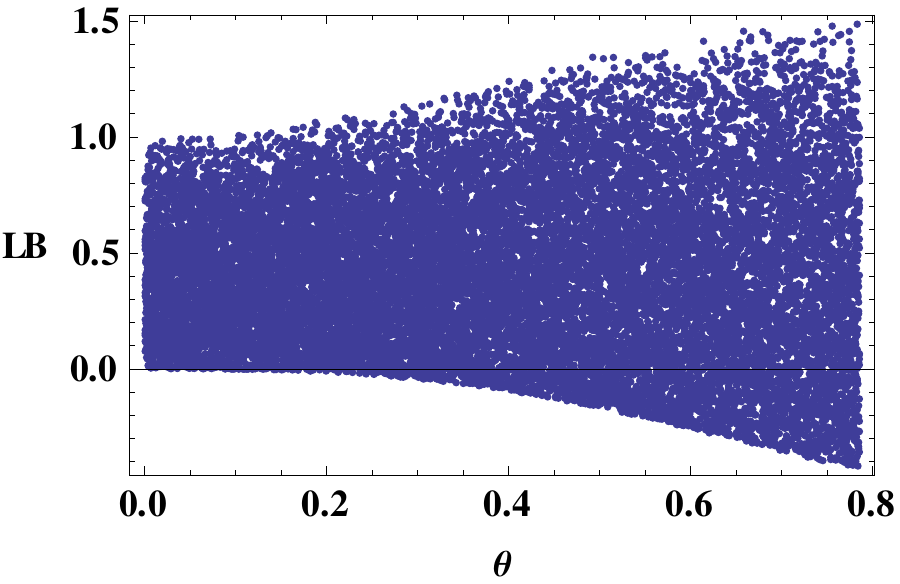}
\includegraphics[width=0.23\textwidth]
{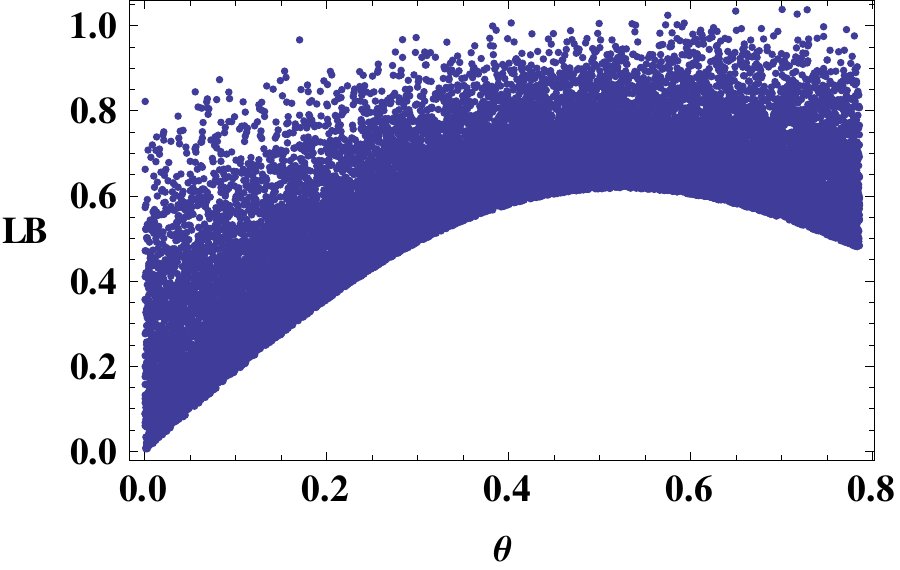}
\includegraphics[width=0.23\textwidth]
{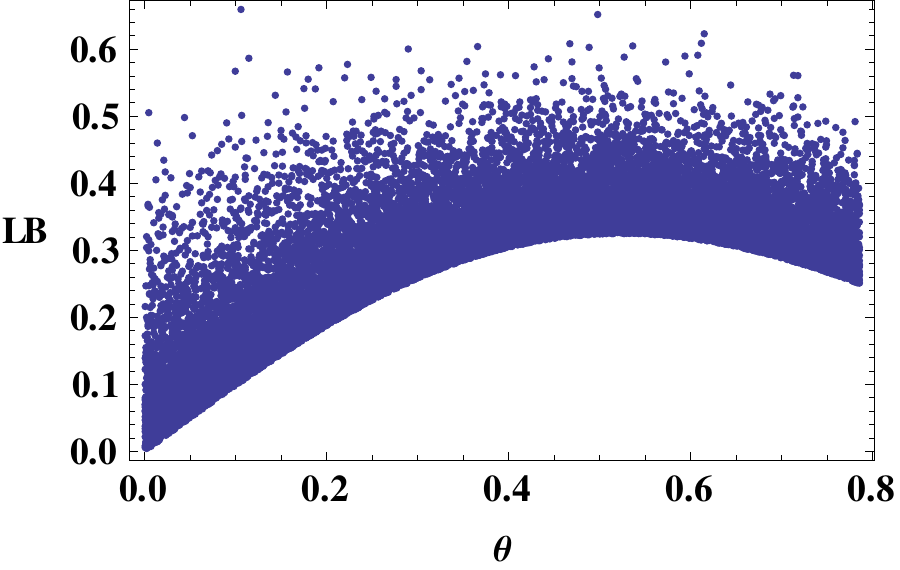}
\includegraphics[width=0.23\textwidth]
{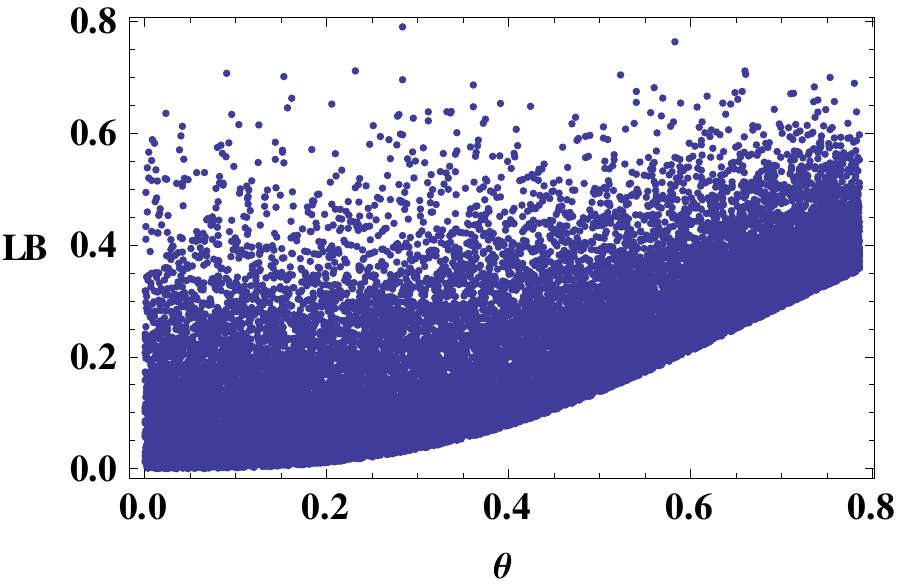}
\caption{\footnotesize{(Color online) Lower bound on the sum of uncertainties of two unitary operators related by the discrete Fourier transform,
given by the random choices of 20 $\Psi^{\perp}$ to $\vert\Psi\rangle=cos\theta\vert 0\rangle -sin\theta\vert d-1\rangle$ for each value of
$0\leq \theta\leq \frac{\pi}{4}$ for dimensions three, five, eight and twelve (starting from top left, in clockwise direction). Here, $d$ denotes the dimension of the Hilbert space.}}
\end{figure}
\vskip 5pt
{\bf Uncertainty relation-3:} The uncertainties of two unitary operators $U$ and $V$ are lower bounded as
\begin{align}\nonumber
\Delta U^2  + \Delta V^2 \geq Max~[&\frac{1}{2}\vert\langle\Psi\vert U+V\vert \Psi^{\perp}_{U+V}\rangle\vert^2, \\ 
    &\frac{1}{2}\vert\langle\Psi\vert U-V\vert \Psi^{\perp}_{U-V}\rangle\vert^2]. 
\end{align}
\vskip 5pt
The proof of this follows from the parallelogram law of the Hilbert space vectors. We take $C=U-\langle U\rangle$  and $D=V-\langle V\rangle$. Then applying the parallelogram law
of vectors, i.e., $2\vert\vert C\vert\Psi\rangle\vert\vert^2+2\vert\vert D\vert\Psi\rangle\vert\vert^2=\vert\vert C+D\vert\Psi\rangle\vert\vert^2+\vert\vert C-D\vert\Psi\rangle\vert\vert^2$ we find, 
$ 2\Delta U^2 +2\Delta V^2= \Delta(U+V)^2 +\Delta(U-V)^2$. It is easily checked that the term $\Delta(U-V)^2$ is positive. Therefore,
for such a case, we have $ 2\Delta U^2 +2\Delta V^2\geq\Delta(U+V)^2$. To obtain Eq(20), we generalize Vaidman's formula \cite{Vaidman1}
$A|\psi\rangle=\langle A\rangle|\psi\rangle+\Delta A|\psi\rangle_{\perp A}$ \cite{Anandan1} for non-Hermitian operator. It can be chekced that for any non-Hermitian operator A, we have 
$A|\psi\rangle=\langle A\rangle|\psi\rangle +\Delta A|\psi\rangle_{\perp A}$, where $\Delta A^2=\langle AA^\dagger\rangle-\langle A\rangle\langle A^\dagger\rangle$. 
If we let $A=U+V$, then $(U+V)|\Psi\rangle=\langle U+V\rangle|\Psi\rangle + \Delta(U+V)|\Psi^\perp_{U+V}\rangle$. 
Thus, we get the above bound. The second inequality follows similarly. In this case also, it is clear that the R.H.S. is indeed a measure of incompatibility of $U$ and $V$ on the state, 
since it expresses the fact that if the incompatible unitary operators do not have a common eigenstate, the R.H.S. always gives a non-zero value showing a finite amount of uncertainty, 
except for the trivial case when $|\Psi\rangle$ is an eigenstate of $U+V$.
The equality condition is again given the Cauchy-Schwartz equality condition. The value for minimum uncertainty states are given in Table I.

\begin{table}
\begin{center}
\begin{tabular}{ |p{0.75cm}|p{1.75cm}|p{1.75cm}|p{1.75cm}|p{1.75cm}|}
\hline
\multicolumn{5}{|c|}{Uncertainty bounds for MUS} \\
\hline
d  & $~~~~\Delta U^2$&UR-1&UR-2&UR-3\\
\hline
2 & 0.5 &0.5&0.5&0.5\\
3 & 0.533494   &0.468072&0.533492&0.454247\\
4 &0.5 &0.416667&0.499721&0.375\\
5 &0.450012&0.370848&0.447758&0.305345\\
6 &0.401089&0.33344&0.39402&0.254505 \\
7 & 0.358678&0.302603&0.348081&0.218263\\
8& 0.323223&0.276769&0.314628&0.191461\\
\hline
\end{tabular}
\caption {Comparison between the minimum uncertainty values given by the uncertainty relations 1, 2 and 3.
Column 1 represents dimension of the Hilbert space. Column 2 represents the the minimum uncertainty of the unitary operators $U$ and $V$, 
which are related by discrete Fourier transform with the constraint $\Delta U^2=\Delta V^2$.
Coulmn-1 is obtained by calculating $\Delta U^2$ in the groundstate of the Harper Hamlitonian (minimum uncertainty state) corresponding to $U$ and $V$. 
UR 1 represents Eq(4), UR 2 represents Eq.[18] and UR 3 represents Eq.[20].}
\end{center}
\end{table}

\section{ V. Minimum uncertainty states }

The minimum uncertainty states are the class of quantum states that minimize the uncertainty functional for the chosen unitary operators \cite{Jackiw}.  
The most straightforward method of finding the minimum uncertainty states have been found by Jackiw, by the \textit{analytic method} \cite{Jackiw}
for the case of the Hermitian operators. Here, one finds the stationary points of the uncertainty functional using
Lagrange's multiplier method. Apart from this method, in the case of two unitary operators related by the discrete Fourier transform [see Eq. (B2) in Appendix B], 
the minimum uncertainty states have been found to be the ground states of a Hermitian operator called the Harper Hamiltonian \cite{Harper}, which is of the form $H= -\cos\theta C_U -\sin\theta C_V$,
where $\theta$ is a real parameter and $C_U$, $C_V$ are Hermitian operators constructed from the operators $C_U=\frac{U+U^\dagger}{2}$ and $C_V=\frac{V+V^\dagger}{2}$, respectively.
The ground states of these Hamiltonian are the discrete generalizations of the coherent states and 
squeezed states in the infinite-dimensional Hilbert space. 
In the infinite-dimensional limit, the Harper Hamiltonian reduces to the Harmonic oscillator Hamiltonian while the discrete coherent and squeezed states reduce to the coherent and squeezed 
states 
in the continuum case of canonical observables \cite{Massar}.

\subsection{A. Minimum uncertainty states as pure states} 

Here, we show that it is enough to consider pure states as the states that minimize the uncertainty
sum as well as the product of the variances. At first, we take the pure state that minimizes the uncertainty functional to be $|\psi\rangle_{min}$. Next,
let us take a density matrix $\rho=\sum_{i=1}^n p_i|\psi_i\rangle\langle\psi_i|$, where, not all of them are the pure states that minimize the product of the variances or the sum of 
the variances. 
Then, we have that $(\Delta U_i)_\rho^2\geq\sum_{j=1}^n p_j(\Delta U_i^2)_{|\psi_j\rangle}$. This inequality follows from the concavity of variance in state for the unitary operators, 
which can be proved easily. 
Thus, we have $\sum_{i=1}^N (\Delta U_i)_\rho^2\geq \sum_{i=1}^N\sum_{j=1}^n p_j(\Delta U_i)_{|\psi_j\rangle}^2=\sum_{j=1}^n p_j\sum_{i=1}^N (\Delta U_i)_{|\psi_j\rangle}^2$. 
But, we know that for the case of the pure states, we have the following equation
\begin{align} 
\sum_{i=1}^N (\Delta U_i)_{|\psi_j\rangle}^2 
\geq \sum_{i=1}^N (\Delta U_i)_{|\psi_{min}\rangle}^2
\end{align}
Thus, from the above equation it is easily seen that for the sum of the variances, among all the states in the state space, only the pure states suffice to desribe all the states that 
minimize
the sum of the variances. Now, let us check this for the product of the variances $\prod_{i=1}^N(\Delta U_i)_{\rho}^2$. For the product of the variances, 
we know that $\prod_{i=1}^N(\Delta U_i)_\rho^2\geq\prod_{i=1}^N(\sum_{k=1}^n p_k(\Delta U_i)_{|\psi_k\rangle}^2)$. Hence, we see that for the minimum uncertainty states, the following 
holds
\begin{align}
\prod_{i=1}^N(\sum_{k=1}^n p_k(\Delta U_i)_{|\psi_k\rangle}^2)=
\sum_{i=1}^np_i^n\prod_{j=1}^N(\Delta U_j^2)_{|\psi_i\rangle}+T
\end{align}
Here, $T$ refers to the remaining cross terms. Since $p_i$ and $\Delta U_i^2$ are all positive, therefore $T$ is itself a positive quantity. Now, by definition, 
$\prod_{j=1}^N(\Delta U_j^2)_{|\psi_i\rangle}\geq\prod_{j=1}^N(\Delta U_j^2)_{|\psi_{min}\rangle}$. As a result, we can replace all $|\psi_i\rangle$ by $|\psi_{min}\rangle$ in $\rho$
to get the minimum value of the product of the uncertainties. Therefore, we get the following equation
\begin{align}
\prod_{i=1}^N(\Delta U_i)_{\rho}^2\geq\prod_{i=1}^N\Delta U_{i}^2)_{|\psi_{min}\rangle}
\end{align}
Thus, we can restrict our search for the minimum uncertainty states to the space of pure states only, which we do in the next section.

\subsection{B. Minimum uncertainty states by \textit{Analytic Method}}

In Ref.\cite{Massar}, the minimum uncertainty states were found out for the product of variances of the unitary operators, 
and lower bounds were found for the uncertainty relation using these states under the constraint
$\Delta U=\Delta V$.
Here, we show that how by using the analytic method of finding the minimum uncertainty states, we recover them as the eigenstates of the Harper Hamiltonian, for the cases of the 
discrete unitary operators
related by the discrete Fourier transform. The analytic method as employed by Jackiw \cite{Jackiw} for Hermitian operators uses the method of Lagrange's multipliers. 
Therefore, applying the same method of Lagrange's multiplier to find the stationary points for the uncertainty
functional of the unitary operators, i.e., the product of the variances of the unitary operators subject to the constraint $\langle\Psi\vert\Psi\rangle=1$, we get the following equation
\begin{align}
 \frac{\delta(\prod_{i=1}^n \Delta U_i^2)}{\delta \langle\Psi\vert}=m\vert\Psi\rangle,
\end{align}
where $m$ is the undetermined multiplier and the variation of $\langle\Psi\vert$ is independent of $\vert\Psi\rangle$. From the above equation, we easily get that
the stationary state equation for the product of the variances of the
$n$ unitary operators subject
to the normalization constraint as the following
\begin{align}
 \frac{1}{n}\sum_{i=1}^n\frac{1}{\Delta U_i^2}\Big[ 1-\frac{(\langle U_i^\dagger\rangle U_i+\langle U_i\rangle U_i^\dagger)}{2}\Big]\vert\Psi\rangle=\vert\Psi\rangle.
\end{align}
We call all the states that satisfy the above equation as {\it{critical states}}. Clearly, every critical state makes the uncertainty product stationary.
This equation has to be solved self-consistently, treating the expectation values constants at first.
The solutions of this equation gives the states that satisfy any generic stationary condition which includes the maxima, minima and the points of inflection.
For the purpose of finding the minimum uncertainty states one has to put the values explicitly and check which particular state gives the minimum value for the product of the 
variances. However, if we connect this to the Harper Hamiltonian, as has been done for the canonical observables where one relates the minimum uncertainty state equation to the 
Harmonic oscillator Hamiltonian, 
we can easily find out the minimum uncertainty state for the chosen unitary operators. For this purpose, we modify equation(13) to match this with the Harper Hamiltonian. 
With $U_1=U$ and $U_2=V$, the equation for the critical states becomes 
\begin{align}
\frac{1}{2}\Bigg[\frac{1-\frac{(\langle U\rangle U^\dagger+\langle U^\dagger\rangle U)}{2}}{\Delta U^2}
+\frac{1-\frac{(\langle V\rangle V^\dagger+\langle V^\dagger\rangle V)}{2}}{\Delta V^2}\Bigg]\vert\Psi\rangle=\vert\Psi\rangle.
\end{align}
Now rearranging the terms we get the following equation
\begin{align}\nonumber 
\Bigg[\Delta V^2 \frac{(\langle U\rangle U^\dagger+\langle U^\dagger\rangle U)}{2}+
\Delta U^2 \frac{(\langle V\rangle V^\dagger+\langle V^\dagger\rangle V)}{2}\Bigg]\vert\Psi\rangle\\  
= [\Delta U^2+\Delta V^2-2\Delta U^2\Delta V^2]\vert\Psi\rangle
\end{align}
On expressing $\langle U\rangle=\vert \langle U\rangle\vert e^{i\phi_U} = \sqrt{ 1-\Delta U^2}e^{i\phi_U}$ and $\langle V\rangle=\vert \langle V\rangle\vert e^{i\phi_V} = \sqrt{ 1-\Delta V^2}e^{i\phi_V}$ 
we have
\begin{align}\nonumber 
\Bigg[\Delta V^2|\langle U\rangle|\frac{(\tilde{U^\dagger}+\tilde{ U})}{2}
+\Delta U^2|\langle V\rangle|\frac{(\tilde{V^\dagger}+\tilde{V})}{2}\Bigg]\vert\Psi\rangle  \\  
= [\Delta U^2\sqrt{(1-\Delta V^2)}\vert \langle V\rangle\vert+\Delta V^2\sqrt{(1-\Delta U^2)}\vert \langle U\rangle\vert]\vert\Psi\rangle.
\end{align}
where, we define $\tilde{U}=e^{-i\phi_U} U$ and $\tilde{V}=e^{-i\phi_V} V $. Also, we define  $C_{\tilde{U}}=\frac{\tilde{U}+\tilde{U}^\dagger}{2}$, $C_{\tilde{V}}=\frac{\tilde{V}+\tilde{V}^\dagger}{2}$. 
Thus $|\langle U\rangle|=|\langle \tilde{U}\rangle|$ and $|\langle V\rangle|=|\langle \tilde{V}\rangle|$. 
Let us take $C_1 =\Delta V^2 \sqrt{(1-\Delta U^2)} $ and $C_2=\Delta U^2 \sqrt{(1-\Delta V^2)}$. Now define
$\frac{C_1}{\sqrt{C_1^2+C_2^2}}= \cos{\theta}$ and $\frac{C_2}{\sqrt{C_1^2+C_2^2}}= \sin{\theta}$.
In this notation, the above equation is simplified to
\begin{align}\nonumber 
[\cos\theta C_{\tilde{U}}
+
\sin\theta C_{\tilde{V}}]\vert\Psi\rangle
= [\cos\theta\vert \langle \tilde{V}\rangle\vert+\Delta \sin\theta\vert \langle \tilde{U}\rangle\vert]\vert\Psi\rangle.
\end{align}

Clearly, the parameter $\theta$ is a squeezing parameter, since for $\Delta U^2 = \Delta V^2$ we get $\theta = \frac{\pi}{4}$, 
which characterizes a coherent state \cite{sch1, Glauber, Sudarshan, Massar} and for $\Delta U^2\neq \Delta V^2$, we get a 
squeezed state \cite{Hall, Massar}. The above equation is the general equation for the critical states for any 
unitary operators $U$ and $V$. The above equation has to be solved self consistently treating the $\phi_U$ and $\phi_V$ constants at first. However, we are mainly
interested in the minimum uncertainty state of the chosen unitary operators. For this purpose, we note that the R.H.S. is independent of $\phi_U$ and $\phi_V$, 
and both $\vert\langle U\rangle\vert $ and $\vert\langle V\rangle\vert $ 
are decreasing functions of the uncertainty functionals $\Delta U$ and $\Delta V$, respectively. Therefore, for a fixed value $\vert\langle U\rangle\vert $ the minimum value of $\Delta V$ will correspond to the
maximum value of $\vert\langle V\rangle\vert$. The same holds for $\Delta U$. Thus, the minimum value of $\Delta U^2\Delta V^2$ will correspond with the maximum value of 
$\cos{\theta}\vert \langle U\rangle\vert+\sin{\theta}\vert \langle V\rangle\vert$ \cite{Opatrny}. Hence, the minimum uncertainty states will be given by the maximum eigenvalue eigenstates of 
$\cos\theta C_{\tilde{U}}+\sin\theta C_{\tilde{V}}$. In other words, the ground states of the operator $\tilde{H}=-\cos{\theta}C_{\tilde{U}}
-\sin{\theta}C_{\tilde{V}} $ will be the minimum uncertainty states. This operator looks similar to the Harper Hamiltonian, 
however a translated version of the Harper Hamiltonian per say. 
It is easy to note that if we can find states such that both $ \langle U\rangle$ and $\langle V\rangle$ are real simultaneously keeping $\Delta U^2\Delta V^2$ invariant, then the minimum uncertainty states 
will be given by the ground states of the Harper Hamiltonian $-\frac{( U^\dagger+ U)}{2}-\frac{( V^\dagger+ V)}{2}$.
Now we illustrate this with the unitary operators obeying the the discrete version of the generalized Clifford algebra at first. It has been proved that for the discrete unitary operators related by the 
discrete Fourier transform, maximum of the $\langle U\rangle$ and $\langle V\rangle$ occurs when $\langle U\rangle = \vert\langle U\rangle\vert e^{-\frac{i2\pi a}{d}}$ \cite{Massar} and 
$\langle V\rangle=\vert\langle V\rangle\vert e^{\frac{-i2\pi b}{d}}$ \cite{Massar}, where $ \{a,b\} = \{0,1,..,d-1\}$ and subsequently 
with the help of the translation operator $U^aV^{-b}$ we can have states for which both $ \langle U\rangle$
and $\langle V\rangle$ are real, keeping the uncertainty product intact. Therefore, we simply find the ground states of the
operator $      
H=-\frac{( U^\dagger+ U)}{2}-\frac{( V^\dagger+ V)}{2}.
$
as the minimum uncertainty states, and other eigenstates as the critical states. This is nothing but the Harper Hamiltonian \cite{Harper, Massar, Opatrny}. It is important point 
to note that we can reduce the above operator in the form of the Harper Hamiltonian only when we can have states for which the average of both $U$ and $V$ are real simultaneously. 
Thus, this method enables one to find the minimum uncertainty state for the unitary operators in general.
The equation presented refers to the stationary states in general, which includes the minimum uncertainty states. An important property of the uncertainty products corresponding to the 
stationary states is that they remain invariant under the action of the translation operators. This is evident from the stationary state equation as well which we show in the next paragraph.

Let us denote a stationary state by $\vert\Psi\rangle_{MUS}$. Let $|\Phi\rangle=U^mV^{-n}|\Psi\rangle$, where $U^mV^{-n}$ is the translation operator, $U$ and $V$ being the unitary operators related by the 
discrete Fourier transform.
It is easy to check that $\Delta U^2$ and $\Delta V^2$ remain invariant under the state transformation of this form. For completeness, we show that the equation for the stationary states also remain invariant 
under this transformation. This is shown by acting the translation operator on the stationary state equation as follows
\begin{align}\nonumber
&\frac{U^mV^{-n}}{2}\Bigg[\frac{1-\frac{(\langle U\rangle_{\Psi} U^\dagger+\langle U^\dagger\rangle_{\Psi} U)}{2}}{\Delta U^2}\\ 
&+\frac{1-\frac{(\langle V\rangle_{\Psi} V^\dagger+\langle V^\dagger\rangle_{\Psi} V)}{2}}
{\Delta V^2}\Bigg](U^mV^{-n})
^\dagger \vert\Phi\rangle= \vert\Phi\rangle.
\end{align}
We modify the above equation by using the commutation relation of the Heisenberg-Weyl operators. As a result, Eq(29) may be expressed as
\begin{align}\nonumber
&\frac{1}{2}\Bigg[\frac{1-\frac{(\langle U\rangle_{\Psi}V^{-n} U^\dagger V^{\dagger -n}+\langle U^\dagger\rangle_{\Psi} V^{-n}UV^{-n\dagger})}{2}}{\Delta U^2} \\
&+\frac{1-\frac{(\langle V\rangle_{\Psi}U^m V^\dagger U^{m\dagger}+\langle V^\dagger\rangle_{\Psi}U^mVU^{m\dagger})}{2}}{\Delta V^2}\Bigg]\vert\Phi\rangle = \vert\Phi\rangle.
\end{align}
By using the commutation relations of $U$ and $V$ we find that
$\langle U\rangle_{\Psi}V^{-n} U^\dagger V^{\dagger -n}\vert\Phi\rangle = \langle U\rangle_{\Phi}U^\dagger\vert\Phi\rangle$.
With this, we have
\begin{align}
\frac{1}{2}\Bigg[\frac{1-\frac{(\langle U\rangle_{\Phi} U^\dagger+\langle U^\dagger\rangle_{\Phi} U)}{2}}{\Delta U^2}
+\frac{1-\frac{(\langle V\rangle_{\Phi} V^\dagger+\langle V^\dagger\rangle_{\Phi} V)}{2}}{\Delta V^2}\Bigg]\vert\Phi\rangle=\vert\Phi\rangle
\end{align}
which is again the stationary state equation satisfied by the states $|\Phi\rangle$.

\section{ VI. Higher dimensional limit of uncertainty relations for two unitary operators related by the discrete Fourier transform}

It has been shown in Ref.\cite{Massar} that there exists a connection 
between the uncertainty relations for the unitary operators and the corresponding Hermitian operators. 
More specifically, in the higher dimensional limit, for a subset of the quantum states, the uncertainty relations of 
the discrete unitary operators related by the discrete Fourier transform, reduce to the Heisenberg uncertainty relations of the corresponding Hermitian operators \cite{Massar}. 
Motivated by this finding, we explore the higher dimensional limits of all the uncertainty relations for the unitary operators. We find that 
these uncertainty relations also reduce to the uncertainty relations of the Hermitian operators for the specific set of states and unitary operators. 
For the first uncertainty relation, higher dimensional analysis shows 
that the uncertainty relation-1 is saturated by the minimum uncertainty states for the case of discrete unitary operators
related by the discrete Fourier transform. 
We prove these results below. Before that, the necessary
condition is that we take only those states that satisfy the relation $\vert\Psi\rangle\in U_\delta(\epsilon)\bigcap V_\delta'(\epsilon')$ for proving the results.

\subsection{A. Asymptotic behaviour of uncertainty relation-1}

For the uncertainty relation-1, we analyze the asymptotic behaviour of Eq(4). Firsr, note that $\Delta U^2+\Delta V^2\simeq\frac{2\pi}{d}(\Delta u^2+\Delta v^2)$. 
The average of the unitary operator $U^{\dagger}V$, when expressed in terms of the Hermitian operators reads as
\begin{align}
\langle U^{\dagger}V\rangle\simeq\langle(I-i\sqrt{\frac{2\pi}{d}}u-\frac{\pi}{d}u^2)(I+i\sqrt{\frac{2\pi}{d}}v-\frac{\pi}{d}v^2)\rangle.
\end{align}
Neglecting the higher order terms, we obtain 
\begin{align} 
\langle U^{\dagger}V\rangle\simeq I+\frac{\pi}{d}(2\langle uv\rangle - \langle v^2\rangle+\langle u^2\rangle) 
+ i\sqrt{\frac{2\pi}{d}}(\langle v\rangle-\langle u\rangle).
\end{align}
Now the other relevant term in the R.H.S. is $ \langle U^\dagger V\rangle\langle U\rangle\langle V^\dagger\rangle+\langle V^\dagger U\rangle\langle V\rangle\langle U^\dagger\rangle$, 
which in terms of the Hermitian operators reads as
\begin{align}\nonumber 
&\langle U^\dagger V\rangle\langle U\rangle\langle V^\dagger\rangle+\langle V^\dagger U\rangle\langle V\rangle\langle U^\dagger\rangle\\ 
&\simeq 2-\frac{4\pi}{d}(\Delta u^2 +\Delta v^2)-\frac{4\pi}{d}\langle u\rangle\langle v\rangle + \frac{2\pi}{d}\langle\{u,v\}\rangle,
\end{align}
wherein we have again only kept terms upto order two. 
Thus, R.H.S. of UR-1 $\simeq\frac{2\pi}{d}(\Delta u^2+\Delta v^2)$, which is nothing but the L.H.S.
This shows that the inequality is saturated for the minimum uncertainty states since they must satisfy the condition 
$\vert\Psi\rangle\in U_\delta(\epsilon)\bigcap V_\delta'(\epsilon')$. From here we infer that the inequality is tight in the case of high dimensions. 
Since we have $\Delta U^2+\Delta V^2\simeq\frac{2\pi}{d}(\Delta u^2+\Delta v^2)$ in the high dimensional limit, as a result we conclude that the inequality is tight in the asymptotic limit 
for the unitary operators also. 

\subsection{B. Higher dimensional limits of other uncertainty relations}

In what follows, we prove that the higher dimensional limit of the uncertainty relation-2, reduce to the stronger uncertainty relation for the Hermitian operators \cite{Maccone}. 
Proceeding in the similar lines as above, i.e., 
by expanding in terms for the Hermitian operators, and keeping terms only upto second order we get L.H.S. as $\frac{2\pi}{d}(\Delta u^2 +\Delta v^2)$. The relevant terms on the R.H.S. are as follows:
\begin{align}\nonumber 
 \vert\langle\Psi^\perp\vert U\pm iV\vert\Psi\rangle\vert^2 \simeq  \nonumber 
 &\vert\langle\Psi^\perp\vert (I+i\sqrt{\frac{2\pi}{d}}u-\frac{\pi}{d}u^2)\vert\Psi\rangle\\ 
 &\pm i\langle\Psi^\perp\vert(I+i\sqrt{\frac{2\pi}{d}}v-\frac{\pi}{d}v^2)\vert\Psi\rangle\vert^2. 
\end{align}
Simplifying and keeping terms upto second order we obtain $\vert\langle\Psi^\perp\vert U\pm iV\vert\Psi\rangle\vert^2 \simeq \frac{2\pi}{d}(\vert\langle\Psi^\perp\vert u\pm iv\vert\Psi\rangle\vert^2)$. Similarly,
we find that
\begin{align} 
 \langle U^\dagger V\rangle-\langle V^\dagger U\rangle\simeq\frac{2\pi}{d}\langle[u,v]\rangle+2i\sqrt{\frac{2\pi}{d}}(\langle v\rangle-\langle u\rangle).
\end{align}
and
\begin{align}
 \langle U^\dagger\rangle\langle V\rangle-\langle V^\dagger\rangle\langle U\rangle\simeq 2i\sqrt{\frac{2\pi}{d}}(\langle v\rangle-\langle u\rangle).
\end{align}
Using the above relations we recover the stronger uncertainty relation for the Hermitian operators $u$ and $v$ as 
\begin{align} 
 \Delta u^2+\Delta v^2\geq \pm i\langle[u,v]\rangle + \vert\langle\Psi^\perp\vert u\pm iv\vert\Psi\rangle\vert^2. 
\end{align}
Thus, from the above equations, we verify that in the higher dimensional limit, the uncertainty relation of the unitary operators give back the stronger uncertainty relation for the Hermitian operators for 
the specified set of the states.
In the same way as above, one can verify easily that writing the unitary operators in terms of the Hermitian operators and keeping the terms upto second order, one is able to recover uncertainty relations of the 
Hermitian operators from the uncertainty relation-3 as well, i.e., we recover the uncertainty relation $\Delta u^2+\Delta v^2\geq \frac{1}{2}\Delta(u+v)^2$, as in Ref.\cite{Maccone}. 
Thus, all the above examples show an interesting link between the uncertainty relation of the unitary operators to that of the Hermitian operators. 

\subsection{C. Higher dimensional limits of the minimum uncertainty states}

Here, we show that, in the high-dimensional the minimum uncertainty state equation for the discrete unitary operators related by the discrete Fourier transform give back the minimum uncertainty state 
equation for the corresponding Hermitian operators.
We rely on the results presented in Ref.\cite{Massar} numerically to bring forth the following observation. We note that minimum uncertainty states belong to 
the set $U_\delta(\epsilon)\bigcap V_\delta'(\epsilon')$. 
Such states have been shown to exist in \cite{Massar}. Thus, the set $U_\delta(\epsilon)\bigcap V_\delta'(\epsilon')$ contains states for which $\Delta U^2$ and $\Delta V^2$ are both 
small. The minimum uncertainty states also have very small values of $\Delta U^2$ and $\Delta V^2$. Thus, the minimum uncertainty states belong to the above set of states.
As a result, we expand the 
unitary 
operators as per Lemma 3, neglect the terms which are very small in the limit of large $d$. We state the observation as follows.
\vskip 10pt
{\bf Observation:} In the high dimensional limit, for small $\delta,~\epsilon$ as defined in section IV, 
the minimum uncertainty state equation for the discrete unitary operators related by the discrete Fourier transform give back the minimum uncertainty state equation for the corresponding Hermitian operators.
\vskip 10pt
First note that, the minimum uncertainty state of the unitary operators $U$ and $V$ are the discretized versions of the
Gaussian functions \cite{Massar}. It has been shown numerically, that these states belong to the particular set.
As a result, we are able to approximate the unitary operators by their series expansion upto the leading order \cite{Massar} as follows
\begin{align}
 U\vert\Psi\rangle\simeq (1-i\sqrt{\frac{2\pi}{d}}+\frac{\pi}{d}u^2)\vert\Psi\rangle.
\end{align}
Similar relation holds for the unitary operator $V$. Then, using the above expression in Eq(26) and 
keeping leading order terms, we get the following equation for the observables $u$ and $v$:
\begin{align}\nonumber
&\frac{1}{2\Delta u^2}\Big[\langle(1+i\sqrt{\frac{2\pi}{d}}u+\frac{\pi}{d}u^2)\rangle(1-i\sqrt{\frac{2\pi}{d}}u+\frac{\pi}{d}u^2)+\\ \nonumber
&\langle(1-i\sqrt{\frac{2\pi}{d}}u+\frac{\pi}{d}u^2)\rangle(1+i\sqrt{\frac{2\pi}{d}}u+\frac{\pi}{d}u^2)\Big]+\\ \nonumber
&\frac{1}{2\Delta v^2}\Big[\langle(1+i\sqrt{\frac{2\pi}{d}}v+\frac{\pi}{d}v^2)\rangle(1-i\sqrt{\frac{2\pi}{d}}v+\frac{\pi}{d}v^2)+\\ 
&\langle(1-i\sqrt{\frac{2\pi}{d}}v+\frac{\pi}{d}v^2)\rangle(1+i\sqrt{\frac{2\pi}{d}}v+\frac{\pi}{d}v^2)\Big]\vert\Psi\rangle\simeq\vert\Psi\rangle.
\end{align}
Now expanding the above equation, and neglecting the higher order terms, we obtain the following equation.
\begin{align}
 \frac{1}{2}[\frac{(u-\langle u\rangle)^2}{\Delta u^2}+\frac{(v-\langle v\rangle)^2}{\Delta v^2}]\vert\Psi\rangle\simeq\vert\Psi\rangle.
\end{align}
This is nothing but the minimum uncertainty state equation of the Hermitian operators as found by the analytical method \cite{Jackiw}. 
This can be understood as that in the infinite-dimensional limit, the 
discretized Gaussian becomes the Gaussian wave functions, i.e., the ground states of the Harmonic oscillator Hamiltonian.

\section{ VII. Conclusions and discussions}

The uncertainty relations are the hallmarks of quantum physics.
To understand and quantitatively capture the essence of preparation uncertainty relations, we have derived several uncertainty relations for the unitary operators.
These uncertainty relations have improved the lower bound of the sum of the uncertainties of the very important class of the discrete unitary operators related by the discrete Fourier transform. 
Apart from this, we have shown numerically that our bound performs better than the existing bound in the current literature. Using the uncertainty relation for the unitary opertars, we have also derived the tight
state independent bound for the two arbitrary Pauli observable.
Furthermore, a higher dimensional analysis of the uncertainty relations in terms of the Hermitian operators reveals that 
they reduce to the stronger uncertainty relations for the corresponding Hermitian operators \cite{Maccone}, for a subset of the quantum states.
We have also shown that the minimum uncertainty state equation of the discrete unitary operators related by the discrete Fourier transform reduce to the minimum uncertainty state equation of the 
corresponding Hermitian operators in the higher dimensional limit. We have prescribed the \textit{analytic method} the minimum uncertainty states of the unitary operators. 
The uncertainty relations are not only a theoretical interesting subject, it has been formulated to be able to observe in the experiments too. In this regard we have given an operational interpretation to the 
uncertainty relations of the unitary operators that brings out preparation uncertainty relations. 
We hope that our results will lead to better understanding of preparation uncertainty using the uncertainty relations for unitary operators and it will be possible to test these relations in experiment.

\section{Acknowledgement}

SB and AP acknowledge financial support from DAE. 
\vskip 25pt
\begin{appendix}
\chapter{\textbf{Appendix A}}

\section{Uncertainty Relation-1 for mixed states}

Uncertainty relation-1 can be generalized to mixed states straightforwardly. For this purpose, we consider the operators
$\overline{U}=U-\langle U\rangle, ~~ \overline{V}=V-\langle V\rangle.\nonumber$
In terms of these operators, we define operator $T=\overline{U}+(\gamma +i\epsilon )\overline{V}$, where $\gamma$ and $\epsilon$ are real parameters. Thus, 
$T^\dagger=\overline{U}^\dagger+(\gamma -i\epsilon )\overline{V}^\dagger$. 
Also, we know that for any operator $T$ \cite{Ballentine}
 \begin{align}
 Tr(\rho ~T~T^\dagger)\geq 0.
 \end{align}
Now to find out $\gamma$ and $\epsilon$ for which $Tr(\rho ~T~T^\dagger)$ is minimum for a given $U$, $V$ and $\rho$, 
we minimize it w.r.t. $\gamma$ and $\epsilon$ and write down the following conditions
\begin{align}
 \frac{\partial}{\partial\gamma}Tr(\rho ~T~T^\dagger)=0, ~ \frac{\partial}{\partial\epsilon}Tr(\rho ~T~T^\dagger)=0
\end{align}
From the above two conditions we get
\begin{align}
 \gamma = \frac{-Tr(\rho(\overline{U}~\overline{V}^\dagger+\overline{V}~\overline{U}^\dagger))}{2Tr(\rho~\overline{V}~\overline{V}^\dagger)}, 
 \epsilon=\frac{iTr(\rho(\overline{U}~\overline{V}^\dagger-\overline{V}~\overline{U}^\dagger))}{2Tr(\rho~\overline{V}~\overline{V}^\dagger)}
\end{align}
We substitute these values in the trace inequality given by Eq(42) and obtain
\begin{align}\nonumber
 Tr(\rho ~\overline{U}~\overline{U}^\dagger).Tr(\rho ~\overline{V}~\overline{V}^\dagger)\geq \\  \frac{1}{4}[(Tr(\rho~(\overline{U}~\overline{V}^\dagger+
 \overline{V}~\overline{U}^\dagger))^2-
 (Tr(\rho~(\overline{U}~\overline{V}^\dagger-
 \overline{V}~\overline{U}^\dagger))^2].
\end{align}
Clearly, $Tr(\rho ~\overline{U}~\overline{U}^\dagger)=\Delta U^2$ and $Tr(\rho ~\overline{V}~\overline{V}^\dagger)=\Delta V^2$. 
Also $Tr(\rho~(\overline{U}~\overline{V}^\dagger+\overline{V}~\overline{U}^\dagger))=\langle UV^\dagger\rangle+\langle VU^\dagger\rangle-\langle U\rangle\langle V^\dagger\rangle-
\langle V\rangle\langle U^\dagger\rangle $. Similarly,
$Tr(\rho~(\overline{U}~\overline{V}^\dagger-\overline{V}~\overline{U}^\dagger))= \langle UV^\dagger\rangle-\langle VU^\dagger\rangle-\langle U\rangle\langle V^\dagger\rangle+ 
\langle V\rangle\langle U^\dagger\rangle$. This gives us 
$\Delta U^2\Delta V^2\geq \frac{1}{4}[(\langle UV^\dagger\rangle+\langle VU^\dagger\rangle-\langle U\rangle\langle V^\dagger\rangle-
\langle V\rangle\langle U^\dagger\rangle)^2-(\langle UV^\dagger\rangle-\langle VU^\dagger\rangle-\langle U\rangle\langle V^\dagger\rangle+ 
\langle V\rangle\langle U^\dagger\rangle)^2]$.
 Rearranging the terms we obtain uncertainty relation-1 for the two arbitrary unitary operators for a general mixed
 state $\rho$ as
 \begin{widetext}
  \begin{align}
  \Delta U^2+\Delta V^2\geq 1+\vert\langle U^\dagger V\rangle\vert^2-\langle U^\dagger V\rangle\langle U\rangle\langle V ^\dagger \rangle-
 \langle V^\dagger U\rangle\langle V\rangle\langle U ^\dagger \rangle.
 \end{align}
 \end{widetext}

\chapter{\textbf{Appendix B}}

 \section{Unitary operators obeying generalized clifford algebra}

The unitary operators
that obey the commutation relations of the generalized Clifford algebra are of special interest, since the infinite dimensional versions of these unitary operators can be written down in terms of 
the Hermitian operators that obey the canonical commutation relations. These unitary operators obey the commutation relation of the form 
\begin{eqnarray}
 UV= e^{i\Phi}VU,~U^\dagger V= e^{-i\Phi}VU^\dagger. 
\end{eqnarray}
The above unitary operators find use in understanding non-local phenomenon in quantum mechanics \cite{Aharonov}. 
The unitary operators $U$ and $V$ can be written as the translation operators $U= e^{\frac{-2ix}{L}}$ and
$V= e^{\frac{-2ip}{P}}$, where the generators of the translation in phase space, i.e., $x(mod~L)$ and $ p(mod~P)$ are called the modular variables\cite{Massar}. In terms of these quantities, the phase can be 
expressed as $\Phi=\frac{4\pi^2}{LP}$.

A particularly important class of unitary operators of the above form is constituted by the discrete unitary operators. 
Among these unitary operators, the ones we consider here, are called the clock and shift matrices \cite{Sylvester, Schwinger}, which are the non-Hermitian generalizations
of the Pauli matrices to higher dimensions. They are the cornerstones in the quantum mechanics of the finite dimensional Hilbert space. The expression for these matrices are given below \cite{Massar}
\begin{align}
U= \sum_{j=-[\frac{d}{2}]}^{[\frac{d-1}{2}]}e^{\frac{i2\pi j\widetilde{k}}{d}}\vert j\rangle\langle j\vert.~~
V= \sum_{\widetilde{k}=-[\frac{d}{2}]}^{[\frac{d-1}{2}]}e^{\frac{i2\pi j\widetilde{k}}{d}}\vert \widetilde{k}\rangle\langle \widetilde{k}\vert. 
\end{align}
The bases $\vert j\rangle$ and $\vert k\rangle$ are the mutually unbiased bases with respect to each other \cite{Ingemar}, and they are related to each other by the discrete Fourier transform as follows
\begin{align}
 \vert j\rangle=\frac{1}{\sqrt{d}}\sum_{\widetilde{k}=-[\frac{d}{2}]}^{[\frac{d-1}{2}]}e^{\frac{i2\pi j\widetilde{k}}{d}}\vert \widetilde{k}\rangle.
\end{align}
The mutually unbiased bases have found wide application in quantum information such as in quantum state determination \cite{r1}, 
quantum state reconstruction \cite{r3}, quantum error correction codes \cite{r4,r5}, detection of quantum entanglement \cite{r6} and the mean King's problem \cite{r7, r8}. Thus, it is important to 
study the uncertainty principle with respect to the mutually unbiased bases, i.e., how much a state can be simultaneously localized in the bases which are mutually unbiased with respect to each other.
From the above relation one can easily see that the two bases satisfy the condition of mutual unbiasedness.
The above matrices, when written in the same bases look like the following 
\begin{align}
U= \sum_{j=-[\frac{d}{2}]}^{[\frac{d-1}{2}]}e^{\frac{i2\pi j\widetilde{k}}{d}}\vert j\rangle\langle j\vert,~~
V= \sum_{\widetilde{k}=-[\frac{d}{2}]}^{[\frac{d-1}{2}]}\vert \widetilde{k+1}\rangle\langle \widetilde{k}\vert.  
\end{align}
The matrices $U$ are called the clock matrices and when they act on the position basis, they just give the eigenstate with an extra phase as the $n^{th}$ root of unity $\omega^n$.
The shift matrices are given by the matrices of the form of $V$. 
The action of the shift matrix on the position eigenbasis produces a shift in the position coordinate by a discrete integer, where the number of shifts is 
given by the number of 
times this matrix is acting on the position basis. It is important to note that the eigenbases of these two matrices are related to each other by the discrete Fourier transform. 
These matrices are also called the Schwinger unitary operators \cite{Schwinger}. 
Since the Schwinger unitary operators are discrete, they obey the integral form of the Heisenberg-Weyl commutation relations \cite{Weyl} 
and form two of the three basis elements of the generalized Clifford algebra \cite{Jagannathan}.
These operators are (up to a phase) the error operators that are used in multidimensional quantum error-correcting codes. 

Since these operators find wide applications in quantum theory, much research has been done on them and some results can be found in Ref.\cite {Massar} that relates
the unitary operators of this form and the corresponding 
Hermitian operators. Following Ref.\cite{Massar}, we express the unitary operators as 
\begin{equation}
 U= e^{i\sqrt{\frac{2\pi}{d}}u}, V= e^{i\sqrt{\frac{2\pi}{d}}v},
\end{equation}
where, $u$ and $v$ are the Hermitian operators. In Ref.\cite{Massar}, the authors have derived a relation between the uncertainty of the unitary operators with the uncertainty relation of 
these corresponding Hermitian operators.
We state their results here, as some of them will be used in our analysis. But, before we state their results, we have to recall some definitions, that are necessary for our proof. 
First we recall the definition of a set $U_\delta(\epsilon)$ as follows.
Let us split the set of indices into two disjoint subsets as $I_{0,\delta}$ and $J_{0,\delta}$.
\begin{align}
 I_{0,\delta}= \{j;\vert j\vert\leq\frac{2}{\pi}[\frac{d}{2}]\delta\},  J_{0,\delta}= \{j;\vert j\vert>\frac{2}{\pi}[\frac{d}{2}]\delta\}, 
\end{align}
where, $\delta>\frac{\pi}{2}$. Then, we have to define a projector $P_\delta=\sum_{k\in I_{0,\delta}}\vert k\rangle\langle k\vert$. Then, the subsets $U_\delta(\epsilon)$ for any $\epsilon>0$
is defined as the set of vectors, such that the following condition is satisfied:
\begin{align}
 \langle\Psi\vert P_\delta\vert\Psi\rangle>1-\epsilon.
\end{align}
With these definitions in hand, Massar-Spindel have proved three important lemmas. 
Now we state the three lemmas as given in Ref.\cite{Massar}, which will be useful for supporting our claims.
Consider a quantum system in a state $|\psi\rangle \in {\cal H}$, where the dimension of the Hilbert space can be finite or infinite. 
\vskip 5pt
\textbf{Lemma 1}:
If $\vert\Psi\rangle=\sum_{j}c_j\vert j\rangle\in U_\delta(\epsilon)$, then $\Delta U^2\leq\frac{\delta^2}{2}+2\epsilon$.
\vskip 5pt
\textbf{Lemma 2}: 
If $\vert\Psi\rangle=\sum_{j}c_j\vert j\rangle\notin U_\delta(\epsilon)$, then $\Delta U^2\leq\frac{\delta^2}{2}+2\epsilon$, then, 
$\exists$ an $\epsilon\leq\frac{(\Delta U^2+\frac{\pi^2}{d^2})}{\sin^2\frac{\delta}{2}}$ and a translation operator $V^k$, such that 
$\vert\Psi\rangle\rightarrow V^k\vert\Psi\rangle\in U_\delta(\epsilon)$.
\vskip 5pt
\textbf{Lemma 3}:
If $\vert\Psi\rangle\in U_\delta(\epsilon)$, then we can expand $U$ as $U\simeq (I +i\sqrt{\frac{2\pi}{d}}u-\frac{\pi^2}{d^2}u^2)$.

We rely on the results presented in their paper numerically to bring forth the following observation. We assume that the set $U_\delta(\epsilon)\bigcap V_\delta'(\epsilon')$ is not null. 
This assumption is natural, as it has been shown to exist numerically \cite{Massar}.
We then notice that there exists at least one minimum uncertainty state (minimum uncertainty state), such that, $\vert\Psi\rangle_{MUS}\in U_\delta(\epsilon)\bigcap V_\delta'(\epsilon')$. 
As a result, we expand the unitary 
operators as per Lemma 3, neglect the terms which are very small in the limit of large $d$. In this limit, it was shown in \cite{Massar} that one has $\Delta U ^2\simeq\frac{2\pi}{d}\Delta u^2$ and 
$\Delta V^2\simeq\frac{2\pi}{d}\Delta v^2$. We use these relations later to analyze the higher-dimensional limits of the uncertainty relations for unitary operators.

The unitary operators obeying the commutation relation of generalized Clifford algebra follow the uncertainty relation as given below \cite{Massar}
\begin{equation}
 (1+2A)\Delta U^2\Delta V^2+A^2(\Delta U^2+\Delta V^2)\geq A^2.
\end{equation}
where, $U$ and $V$ are discrete unitary operators obeying the commutation relation $UV= e^{i\Phi}VU$ and $U^\dagger V= e^{-i\Phi}VU^\dagger$ and $A= \tan{\frac{\Phi}{2}}$. This uncertainty relation is 
interesting since it interpolates between the uncertainty of the Pauli sigma matrices in $d=2$ limit, and the Heisenberg uncertainty relation for the Hermitian operators (corresponding to the unitary operators) 
in the infinite-dimensional limit. Also, it provides an uncertainty relation of the modular variables, finding importance in understanding the non-local phenomenon \cite{Aharonov}.

\end{appendix}

\end{document}